\documentclass{aa}
\usepackage{graphicx}
\usepackage{txfonts}

\begin{document}

%DC Changed acknowledgement 30/1/2006 - included GO programme number, deleted the Archive reference
%DC This is very important, please leave this!!!

\title{Globular cluster systems of six shell galaxies
	\thanks{Based on observations made with the NASA/ESA Hubble Space Telescope, 
obtained at the Space Telescope Science Institute, 
which is operated by the Association of Universities for Research in Astronomy, 
Inc., under NASA contract NAS 5-26555. These observations are associated with program GO9399
}}
%DC - order of authors changed as proposed by Reynier and agreed
\author{G.Sikkema\inst{1} \and R.F. Peletier\inst{1} \and D.Carter\inst{2}
\and E.A. Valentijn\inst{1} \and M. Balcells\inst{3}  }

\institute{Kapteyn Astronomical Institute, University of Groningen, PO Box 800, 9700 
AV Groningen, The Netherlands \and Astrophysics Research Institute, 
Liverpool John Moores University, 12 Quays House, Egerton Wharf, Birkenhead, 
CH41 1LD, United Kingdom \and Instituto de Astrof\'isica de Canarias, V\'ia 
L\'actea s/n, La Laguna, 38200, Spain}

\date{}
%Received $\<$date$\>$ / Accepted $\<$date$\>$}

%DC Abstract rewritten by DC 30/1/2006     

\abstract
{Shells in Elliptical Galaxies are faint, sharp-edged features, believed to provide evidence of a recent 
($\sim 0.5 - 2 \times 10^9$ years ago)
merger event. We analyse the Globular Cluster (GC) systems of six shell elliptical
galaxies, to examine the effects of mergers upon the GC formation history.
}
{We examine the colour distributions, and investigate differences between  
red and blue globular cluster populations. We present luminosity functions, 
spatial distributions and specific frequencies ($S_N$) at 50 kpc radius for our sample.
}
{We present V and I magnitudes for cluster candidates measured with the HST Advanced Camera for 
Surveys (ACS). Galaxy background light is modelled and removed, and magnitudes are measured in 8 
pixel (0.4 arcsec) diameter apertures. Background contamination is removed using counts from Hubble Deep Field 
South.
}
{
We find that the colour distributions for NGC 3923 and NGC 5982 have a
bimodal form typical of bright ellipticals, with peaks near $V-I=0.92 \pm 0.04$ and
$V-I=1.18 \pm 0.06$. In NGC 7626, we find in addition a population of abnormally luminous
clusters at $M_I=-12.5$. In NGC 2865 we find an unusually blue population, which may also be young.
In NGC1344 and NGC474 the red cluster population is marginally detected.
The radial surface density profiles are more flattened than the galaxy light in the cores.
 As already known, in NGC3923, which has a high $S_N$ of 5.6, the radial density distribution is more shallower than the diffuse galaxy light.
}
%NEW
{The clusters in NGC 2865 and NGC 7626 provide evidence for formation of a population associated with a recent 
merger. In the other galaxies, the properties of the clusters are similar to those observed in other, non-shell, elliptical galaxies.
}

%DC Keywords chosen by DC 30/01/2006
\maketitle

\newpage
\ \\

%DC minor rewording throughout this section 30/1/2006
\section {Introduction}

In current galaxy formation models, most ellipticals have already 
formed at $z > 2$ (\cite{ellis}; \cite{peeb2}; \cite{d2}). 
It is unclear whether all globular cluster (GC) systems were also formed 
at this early epoch or if substantial numbers are still forming today. An important 
diagnostic is the existence of bimodality in the colour distribution of GCs, present 
in many early type galaxies (\cite{ashzepf2}, \cite{whitmore95}). Generally this is 
explained as being due to metallicity differences indicating two or more populations 
of GCs. Several theories have been proposed to explain the origin of bimodality:

Merger scenarios (\cite{Toomre}; \cite{Schweizer87}; \cite{ashzepf}) in which the 
metal-rich GCs were created in gas rich mergers. Since most star formation occurred
at early epochs, this means in general that the metal-rich GCs are also old. However, 
this scenario also suggests that GCs can still be forming today in mergers. This is 
supported by observations of young cluster-like objects in current mergers in action 
or possible merger remnants like NGC 4038/39 (\cite{whitschw}; \cite{whitm}), 
NGC 3921 (\cite{schweiz}), NGC 7252 (\cite{miller}), NGC 1316 (\cite{goudfrooij}; 
\cite{goudfrooij3}; \cite{goudfrooij2}), NGC 1700 and NGC 3610 (\cite{whit97}).

Two others models explaining bimodality are the accretion model (\cite{cote}) and 
the multiphase formation model (\cite{forbesbrod}; Harris et al. 1998). In both 
models all GCs are old. The first model produces bimodality by accreting and mixing 
metal poor GCs from dwarf galaxies with the more metal rich GCs of the massive host 
galaxy. Cannibalism by our own galaxy of the Sagittarius and Canis Major dwarf galaxies 
and their clusters (\cite{ibata}; \cite{forbstra}) and observations of large numbers of 
dwarf galaxies around giant galaxies are cited as supporting this scenario.
The second model explains the bimodality as the result of two phases of GC formation in 
the initial collapse and formation of a galaxy. The metal poor clusters, and a small proportion of
the stars form in the initial gravitational collapse, then the metal rich clusters
and the bulk of the stellar component form from enriched gas in a second collapse phase
about one or two Gyr later. \cite{strader2} argue that this ``in situ'' model of GC formation
is in better agreement with their observations of the correlation of the colours
of the metal-poor populations with galaxy luminosity.

%DC - above - changed to better reflect what the Strader at al reference really is saying

A combination of these different scenarios is used in the hierarchical merging model of 
\cite{beasley}, who undertook semi-analytical simulations of GC formation. In this model 
the metal-poor GCs are old and formed in cold gas clumps, the metal-rich ones are created 
later in merger events. In the hierarchical build up of galaxies, accretion of GCs 
will also take place. These simulations are able to reproduce the many variations in the 
colour distributions of GC systems observed in elliptical galaxies as well as the 
observed $L - N_{tot}$ relation.

%NEW
Recently, Yoon et al. (2006), showed that the apparent bimodality in 
globular cluster colours not necessarily implies a
bimodal metallicity distribution. The nonlinear nature of the 
metallicity-to-colour  transformation could cause a single old
population with a unimodal metallicity distribution to look bimodal.  This 
model is attractive because it gives a very simple
explanation for the observed distributions and could simplify theories of 
elliptical galaxy formation. However, the
observations of recent GC formation, as mentioned above will likely 
sometimes disturb the predictions made by this model.

%DC - changed following section to include more references as requested by referee
%DC - also to be clearer about age differences, because the merger GCs do not have to
%DC - be young

There are many examples of multi-colour and spectroscopic data for the GC
systems of 'normal' elliptical galaxies, and these generally give old ages
for both blue and red populations (M49: \cite{puzia}, \cite{cohen}; NGC 1399: \cite{forbesbeas}; M87: 
\cite{cohen2}; NGC 1052 and NGC 7332: \cite{forbes}; and in a sample of early-type
galaxies: \cite{strader3}). This is in contrast
to studies of the well-known bimodality of the colour distribution of the GCs
of the LMC (\cite{gk}; van den Bergh 1981, 1991) which is
primarily an age effect, understood in terms to the evolution with time of
clusters in the colour-colour diagram (\cite{ff}), possibly combined
with a relationship between age and metallicity (\cite{bc}; \cite{girardi}). In ellipticals, 
the majority of GCs formed at
high redshift (z$>$2.5), whichever formation mechanism is dominant. However it is important 
to examine evidence that {\sl recent} merger events can produce enhanced populations of
young clusters, as this process may have been much more important in the early universe.

%DC More material added here as the referee has asked for an introduction to shells

This study focused on a sample of elliptical galaxies with faint stellar shells in their envelopes. 
Shells and ripples in elliptical galaxies (\cite{malin80}; \cite{malincarter};
\cite{schweizseitz}) are faint, sharp edged stellar features in the envelopes of these
galaxies which are the remnants of the stellar components of minor mergers in the 
comparatively recent history of the galaxy (\cite{quinn}). Typical 
dynamical ages of the shell systems are $\sim 0.5 - 2 \times 10^9$ years (\cite{nulsen};
\cite{hernquist}). If
we can detect a population of clusters then the age of this population will provide an 
independent estimate of the age since the merger, assuming that the same event
produced both the shells and the young clusters. In NGC 2865, one of our sample,
Hau et al. (1999) find that the age of a nuclear starburst model for the young stellar population
in the core of this galaxy is much older than the dynamical age of the shells, although 
they do find a better correspondence between ages for a model involving truncation of
ongoing star formation. In NGC 1316, Goudfrooij et al. (2001b) find a cluster system with an age
of $3.0 \pm 0.5$ Gyr, consistent with the age of the nuclear stellar population. The 
shell systems of both NGC 2865
and NGC 1316 are complex, Type II or III systems (\cite{prieur90}) whereas the dynamical 
age estimates are more directly applicable to simple phase wrapped, Type I shell systems.

%DC - Substantial rewrite as requested by referee, incorporating older references
%OLD
%Two techniques for investigating the age differences between different GC populations
%NEW minor rewording
Two techniques to investigate possible age differences between different GC populations
involve measuring radial density profiles, and globular cluster luminosity functions
(GCLFs). Radial density profiles generally show a flattening of the globular cluster
density profile near the centre, when compared with the profile of the background
galaxy light (\cite{lauer}; \cite{donna}). Mechanisms which could cause a depletion 
of the cluster population near the centre are dynamical friction, which causes clusters
to spiral in towards the centre (\cite{tremaine}, \cite{pesce}), and destruction by tidal shocks as the
clusters pass close to the nucleus (\cite{ostrikerbinney}; \cite{ct}). This second 
process operates preferentially in triaxial potentials, and on clusters on radial orbits. 
If we can identify a population of younger clusters, created during a recent merger, then 
the density profile might extend further into the centre as the clusters have had less 
time to disrupt. However, this does depend upon the orbital structures to be the same, 
if one or other population were on predominantly radial orbits, then this would cause
a stronger flattening of the core. 

The same process can lead to evolution of the mass function of the clusters and hence 
the GCLF (\cite{fallrees}, \cite{gnedin}, \cite{fallzhang}). Lower mass clusters
are preferentially destroyed, leading to the well known turnover in the GCLF. 
A younger population would contain more low-mass clusters, and the GCLF should
be closer to the original mass function, which might be a power law. Observations
of young cluster systems such as NGC 1316 (\cite{goudfrooij3})
and NGC 7252 (\cite{miller}) do show power law GCLFs.

%DC - minor rewording below

In this paper we analyse the properties of the GC systems of six shell 
elliptical galaxies using optical V and I data from the Advanced Camera for Surveys (ACS) on HST. 
Although the observations were optimised to study the 
shell structures rather than the GCs, the magnitudes, colours and spatial distribution
will indicate whether GC systems of any of the shell galaxies differ from 
those of normal early type galaxies. 
We will also provide data on three GC systems which have not been studied before: 
NGC 474, NGC 1344 and NGC 2865. The last galaxy is known as a recent merger remnant 
(\cite{hau}; \cite{schimi}). The GC systems of three of our galaxies have been studied 
before (NGC 3923, ground-based (\cite{zepf94}; \cite{zepf95}); NGC 5982 and NGC 7626, WFPC2 
on the HST (\cite{forbesfranx}). The larger field of view and higher resolution of ACS 
will provide more detections and more complete knowledge of these GC systems.

In Section 2 and 3 the observations and data reduction are described, which include the 
detection and selection of the GCs, calculation of completeness levels and photometry of 
the globular clusters. Section 4 presents the V-I colour distributions and spatial 
distributions of the globular cluster systems. Section 5 describes the globular cluster 
luminosity function which is used in Section 6, where the specific frequencies are 
calculated. Finally in Sections 7 and 8 we discuss the results and present the conclusions.
 
\section {Observations and Data Reduction}
\label{sec:observations}

\begin{table*}
\centering
%\vspace{6cm}
%{\bf Table 1}: Properties of six shell galaxies \\
\begin{tabular} {l r r r r r r r r l l l l}
\hline
\hline

Galaxy&	RA (J2000)&  DEC(J2000) &l &b & $A_V$ & $A_I$ & type & $m_V$   & m-M & d  & Dn-$\sigma$ & $\sigma$  \\
\hline
    &           &          &          &           &  mag   &   mag     &       &    mag     &  mag &   Mpc    &  Mpc   & km/s \\
\hline
(1) &    (2)    &   (3)    &    (4)   &   (5)     &   (6)  &   (7)     &  (8)  &    (9)     &  (10)&   (11)   &  (12)  &  (13)\\
\hline
\object{NGC 474}&  $1^h 20^m 06^s.7$  & $+03^\circ 24' 55''$ & 136.80  & $-58.68$  & 0.11& 0.07&E?  & 11.39 & 32.56 $\dagger$ & 32.5   &  - &164\\
\object{NGC 1344}& $3^h 28^m 19^s.7$  & $-31^\circ 04' 05''$ & 229.07  & $-55.68$  & 0.06& 0.04&E5  & 10.41 & 31.48  & 22.1 $\ddagger$ & 20 &187\\
\object{NGC 2865}& $9^h 23^m30^s.2$   & $-23^\circ 09' 41''$ & 252.95  & $+18.94$  & 0.27& 0.16&E3-4& 11.30 & 32.89  & 37.8            & 25$\diamond$ &230$\diamond$ \\
\object{NGC 3923}& $11^h 51^m01^s.8$  & $-28^\circ 48' 22''$ & 286.53  & $+33.32$  & 0.27& 0.16&E4-5& 9.88  & 31.80  & 20.0 $\ddagger$ & 21 &249\\
\object{NGC 5982}& $15^h 38^m39^s.8$  & $+59^\circ 21' 21''$ &  93.10  & $+46.92$  & 0.06& 0.04&E3  & 11.20 & 33.11 $\dagger$ & 41.9   & 41 &240\\
\object{NGC 7626}& $23^h 20^m 42^s.3$ & $+08^\circ 13' 02''$ &  87.86  & $-48.38$  & 0.24& 0.14&Epec& 11.25 & 33.41 $\dagger$ & 48.2   & 46 &270\\

\hline
\label{info}
\end{tabular}
\caption{Properties of six shell galaxies. Data in columns 2-9 from Roberts (1991). 
1st column: Galaxy name; 2nd and 3rd column: Right Ascension and Declination;
4th and 5th column: galactic longitude and latitude; 6th and 7th column: 
extinction coefficients in V and I magnitudes from Schlegel et al. (1998); 8th 
column: Morphological type (throughout this paper we assumed that NGC 474 is 
elliptical and not S0 (\cite{hau2}); 9th column: total apparent V magnitude; 10th column: SBF 
distance modulus in I, corrected for extinction. (\cite{tonry}). $\dagger$: 
distance moduli taken from \cite{roberts} determined from HI velocity data, 
corrected for galactic rotation and restframe of the Local Group using $H_0$=75km/s/Mpc. 
11th column: distances adopted in this paper using column 10 except $\ddagger$: 
GCLF distances for NGC 1344 
and NGC 3923, calculated in Section 5.2.; 12th column: Dn-$\sigma$ distances from Faber et al. (1989); 13th column Central velocity dispersion 
$\sigma$; typical errors are 10 km/s (HYPERLEDA$^4$)
$\diamond$: d(Dn-$\sigma)$ for 
NGC 2865 is probably wrong, since NGC 2865 exhibits a central depression in $\sigma$ 
due to a rotating disk (\cite{hau}). Extrapolating his data gives $\sigma$=230km/s 
(d(Dn-$\sigma$)=34.7Mpc), much larger than the value used by Faber et al. ($\sigma$=168km/s).}
\end{table*}

The six shell galaxies (see Table 1) were observed with the ACS\_WFC camera between 
July 2002 and January 2003 with the filters F606W (V) and F814W (I), with CR\_SPLIT=2. 
The ACS camera contains two CCDs of 2048 x 4096 pixels, each pixel having a size of 
$0.049''$ pixel$^{-1}$ resulting in a field of view of $202''$ x $202''$. Exposure 
times were on average 1000s, see Table 2. The inner 8 pixels of NGC 474  and the 
inner 24 pixels of NGC 2865 were saturated in both V and I. 

%DC - very minor rewording below

Standard reduction was carried out in the IRAF+STSDAS\footnote{IRAF is 
distributed by the National Optical Astronomy Observatories, which are operated by 
the Association of Universities for Research in Astronomy, Inc., under cooperative 
agreement with the National Science Foundation.} environment, using the packages 
CALACS and PyDrizzle. These are provided by the Space Telescope Science Institute 
(STScI). CALACS processing includes bias and dark subtraction, removal of the 
overscan regions, flat fielding and cosmic ray rejection. The default pipeline is 
not efficient at removing cosmic rays when the image is filled by a large galaxy. 
To solve this problem we changed the value of CALACS pipeline parameter SCALENSE from 0.3 
to 0.0. Setting SCALENSE to 0.0 increases the probability of removing good stellar 
data\footnote{http://www.stsci.edu/hst/acs/documents/newsletters/stan0301.html}. 
However, this is only true for empty fields, which is not the case for our data. 
The resulting images were further processed by PyDrizzle, which removes the geometric 
distortion of the ACS optical configuration. Finally, after drizzling the images, the 
IRAF package LA\_COSMIC (\cite{dokkum}) was used to remove any remaining cosmic rays, 
which still affected several hundreds of pixels in each image. The standard ACS photometric 
calibration was used (\cite{sirianni}) to obtain Johnson V and Cousins I magnitudes. The following 
transformation formulae were applied:

\begin{equation}
V_{J}=m(F606W)+26.331+0.340*(V-I)_{JC}-0.038*(V-I)_{JC}^2
\end{equation}
\begin{equation}
I_{C}=m(F814W)+25.496-0.014*(V-I)_{JC}+0.015*(V-I)_{JC}^2
\end{equation}

Here m(F606W) and m(F814W) are ACS Vega instrumental magnitudes and V and I are in 
Johnson and Cousins systems respectively. 
The Johnson and Cousins magnitudes were corrected for galactic extinction using the values 
of Schlegel et al. (1998), which are listed in Table 1. 

\section{Data Analysis}
In this Section we describe how the globular cluster source catalogues were obtained and 
describe their characteristics in terms of photometric errors and completeness, which are 
used in the further analysis of the data. 
% We produced galaxy-subtracted images, using 
%the GALPHOT software (see \cite{jorg}). Source detection and photometry was done with 
%SExtractor. 

\subsection{GALPHOT}
Information about the morphology of the galaxies was obtained by using the ellipse fitting 
task GALPHOT (see \cite{jorg}); it returns information such as ellipticity, position angle, 
surface brightness and the C3,C4,S3,S4 coefficients (\cite{carter80}), all as a function 
of radius. A galaxy subtracted residual image is also returned, which we used to extract 
the GCs.

In the GALPHOT processing, background galaxies, point-like objects, dust lanes, bright 
pronounced shells and additional bad data were masked out in an iterative way. Remaining 
faint shell structures, having a brightness typically not more than $5\%$ of the galaxy 
light, do not severely affect the final results. The best fits were obtained by allowing the 
center, position angle and ellipticity to be free parameters. In two cases: NGC 2865 and 
NGC 5982 the central regions could not be subtracted in a proper way, these regions 
correspond to circles with diameters  $4''$ for NGC 2865 and $24''$ for NGC 5982 
respectively. Note that the inner $0.3''$ is saturated in the center of the latter galaxy.

%DC - minor rewording

Light profiles were obtained by plotting the surface brightness for each fitted ellipse 
as function of radius. The outer parts of the light profiles are severely affected by 
uncertainty in the determination of the 
background. It is difficult to determine a reliable background values from the ACS images 
themselves, since the galaxies fill the whole field. Fortunately, for five galaxies we 
found optical wide field data in the R band in the ESO archive of WFI@2.2m. The WFI 
camera has a field of view of $34'$ x $33'$, much larger than the galaxies. 

%to reduce the WFI images.  and we could independently determine the background.
%which causes problems if colours maps are made and if outer shell colours are to be compared 
%with outer galaxy colours. It is difficult to determine the background  

\begin{table*}[]
\begin{tabular} {l c c c c c c }
\hline
\hline
(1) &    (2)    &   (3)    &    (4)   &   (5)     &   (6)  &   (7)   \\
\hline
Galaxy & exp. V (s)& exp. I (s) & 80\%V (mag)  & 80\%I (mag) & bg. V (cnts)& bg. I (cnts) \\ %& p1 & amp1 & fw1& p2& amp2 & fw2\\
\hline
NGC 474 & 1140& 960& 25.74 &  24.78&203 & 129 \\       %& 0.72 & 46  & 0.25 &  -   &   - &  -  \\
NGC 1344& 1062& 840& 25.81 &  24.76& 94 &  47  \\       %& 0.72 & 70  & 0.24 &  -   &  -  &  -  \\
NGC 2865& 1020& 840& 25.85 &  24.78&146 & 109 \\       %&  -   & -   &-     &  -   &-    & -   \\
NGC 3923& 1140& 978& 25.59 &  24.47&170 & 103 \\       %& 0.66 & 110 & 0.14 & 0.84 & 132 & 0.17\\
NGC 5982& 1314&1020& 26.08 &  24.80&124 & 182 \\       %& 0.73 & 65  & 0.13 & 0.91 & 65  & 0.18\\
NGC 7626& 1140& 960& 25.96 &  24.86&119 &  80  \\       %& 0.67 & 119 & 0.15 & 0.85 & 147 & 0.16 \\
\hline
\label{compTabel}
\end{tabular}
\caption{Observational characteristics. 1st column galaxy name, 2st and 3nd column: 
V and I exposure times in seconds, 4th and 5th column: 80\% completeness levels in 
V and I, 6th and 7th column: adopted background value in counts}
\end{table*}

We used the ASTRO-WISE system \footnote{www.astro-wise.org/portal} (\cite{vk}) to reduce the WFI images. After subtracting a 
constant background value from the WFI images, GALPHOT was applied to them. 
The ACS background values were determined by matching ACS light profiles to the WFI light 
profiles. For the galaxy without WFI data, NGC 5982, we extrapolated from the outer 
points of the image using a \cite{vau} $r^{1/4}$ law. The calculated 
values for the backgrounds are listed in Table 2.
The final GALPHOT results describing the morphology of the six galaxies will be presented in 
a future paper (Sikkema et al. in preparation). 

\subsection{Globular cluster candidates}
The galaxy subtracted, residual images were used for source detection and photometry of 
globular cluster candidates (GCCs). SExtractor (\cite{bertin}) was used for this purpose. 
The keyword BACK\_FILTERSIZE was set to 5; the keyword BACK\_SIZE was set to the relatively 
small value of 24, to account for background variations caused by broad shell regions, 
diffracted light from bright stars, and dusty regions. 
\begin{figure}[!h]
    \resizebox{\hsize}{!}{\includegraphics{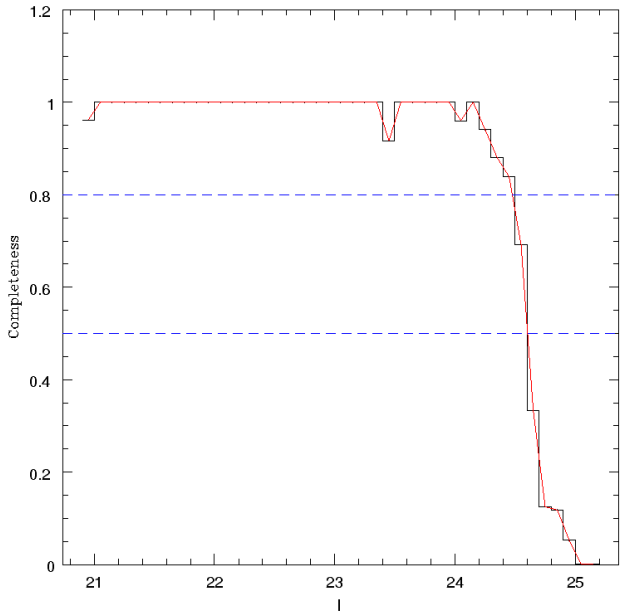}}
    \caption{Histogram completeness for NGC 3923 I band image, using the gc-type object. 
Vertically: completeness ratio; horizontal scale: instrumental ACS magnitudes. Dashed lines 
indicate 80\% and 50\% completeness levels.}
    \label{completeness}
\end{figure}
\footnotetext[4]{http://www-obs.univ-lyon1.fr/hypercat}

\subsection{Photometry}
\label{sec:photometry}

Because we are analysing point like objects, the magnitudes are measured in fixed apertures,
and we tested
apertures with diameters of 4,6,8,10,12,14,16,18,20 pixels. The quality of the 
photometry was analysed by adding objects to the residual images using the IRAF task 
MKOBJECT (the object was extracted from an ACS image and is a typical PSF object with 
a few thousands counts at its peak). We did this 5 times, each time introducing 350 
objects with a uniform magnitude distribution and positions (generated by IRAF/STARLIST. 
Catalogues were extracted by using SExtractor again with the same parameters as before. 
The resulting catalogues in V and I were then associated, keeping only associations within 
positional error ellipses of 1xFWHM=2.4 pixels. Comparing input STARLIST data with output 
data gives information about photometric errors and completeness levels.

\subsection{Completeness}

We define completeness as the ratio of recovered objects divided by the original 
number of added objects measured in a particular magnitude bin. Figure \ref{completeness} 
shows a typical result of this procedure for the I band residual image of NGC 3923. On 
the horizontal axis are the input STARLIST magnitudes. The completeness is plotted 
on the vertical axis. The horizontal lines indicate 50\% and 80\% completeness levels. The results 
for each image are shown in Table 2.
%One complication is that the completeness levels depend on the type of object. 

\subsection{Photometric errors and aperture selection}

\begin{figure}[!h]
    \resizebox{\hsize}{!}{\includegraphics{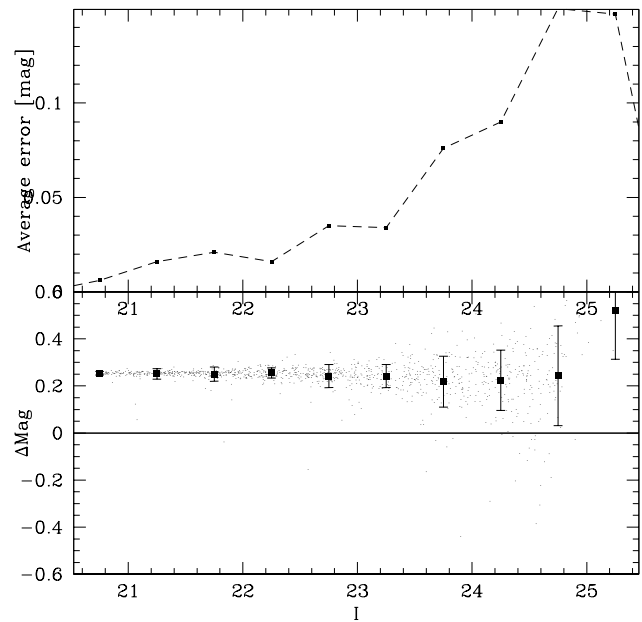}}
    \caption{Photometric errors for artificial pointlike objects in the GALPHOT NGC 3923 
I band residual image. {\bf{Top:}} RMS errors in magnitudes in each bin; 
{\bf{bottom:}} Off-sets from STARLIST artificial objects using an aperture of 8 pixels.}
    \label{photErrors}
\end{figure}

The photometric errors were obtained by subtracting input STARLIST magnitudes from output 
aperture magnitudes, where extreme outliers (mainly due to false associations) in each 
magnitude bin were removed using 5x sigma clipping. Typical results, using the NGC 3923 I band 
residual image and an 8 pixel aperture, are shown in Figure \ref{photErrors}. The upper 
panel shows RMS errors in magnitudes in each bin; off-sets from STARLIST magnitudes are 
plotted in the lower panel. The magnitude errors depend on the different apertures in 
various ways:

\begin{itemize}
\item  The rms-errors at the 80\% completeness levels increase for larger apertures: 
from 0.08 mag using a 4 pixel aperture, to 0.18 mag using a 20 pixel aperture.
\item  The measured magnitudes show systematic off-sets for each aperture; larger 
apertures give smaller offsets. We assume that the offsets of the real data are the same in all of our twelve images. 
\item  The off-sets vary within each magnitude bin; however the variations are larger for large apertures. 
\end{itemize}

Considering these three error sources leads to a optimum aperture diameter of 8 pixels, which 
was used in all remaining analysis. The photometric error is $\approx 0.10$ mag. near the 
80\% completeness levels and the aperture correction is 0.26 mag. Figure \ref{photErrors} 
shows an example using this aperture for the NGC 3923 I band residual image.

Before doing the photometry, we checked whether globular clusters are resolved in our 
galaxies. Table 1 shows that our closest galaxies, NGC 1344 and NGC 3923, have a distance 
of about 20 Mpc. At this distance, typical galactic globular clusters, which have a 
half total light diameter of 6 pc (\cite{vdb}), will have an angular size of somewhat more than 
1 ACS pixel. We conclude that most globular clusters will be unresolved.

\subsection{Selection of GCCs}
\label{sec:selection}

\begin{figure}
    \resizebox{\hsize}{!}{\includegraphics{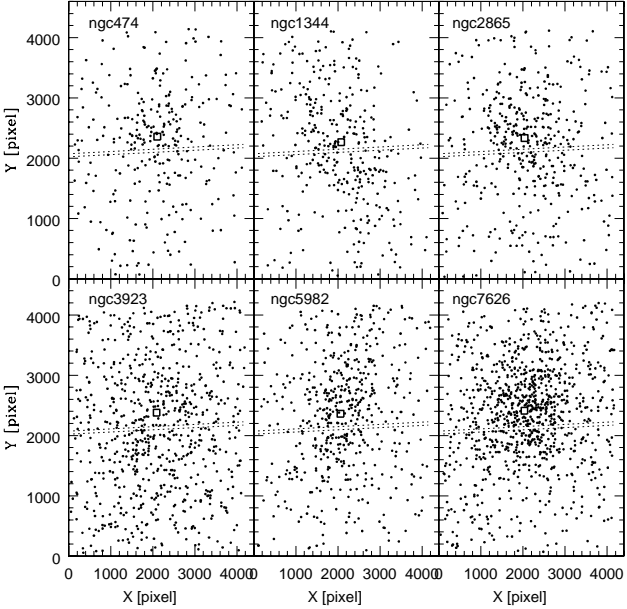}}
    \caption{Location of globular clusters, using ACS pixel coordinates (pixelsize = 
0.05 arcsec; the open square represents the center of the galaxy.}
    \label{LOC}
\end{figure}

We made use of various SExtractor keywords to select the globular cluster candidates. 
We kept all objects which have ELONGATION $< 1.4$ (keep round looking objects), 
$ 1.8 <$ FWHM\_IMAGE $< 5$ (removing most spurious objects and extended objects) and 
FLAGS $> 0$ (removing objects which are blended, saturated or placed near other bad pixels).
Objects fainter than the 80\% completeness levels were also removed. The remaining lists 
of objects in V and I were associated using positional error ellipses of $1\times$FWHM=2.4 
pixels, which again removed many objects. The final source list was further cleaned by visual 
inspection of the residual images, thereby removing any false data, for instance bright 
galaxy cores, sources located near borders or detections in spikes of bright stars. 
Figure \ref{LOC} shows the distribution of the remaining objects.

Before using these data, we note that the source catalogue is still contaminated 
by two sources: foreground stars and background galaxies. Below, an estimate 
of these numbers is made.

\begin{enumerate}
\item Galactic foreground stars. The number of foreground stars depends upon galactic 
coordinates. Table 1 shows that NGC 2865, with the lowest galactic latitude of $19^{\circ}$, 
is likely to be affected most by foreground stars. We estimated the number of stars in each 
field by using The Besan\c con model (\cite{robin}). We used the following input for the model:

\begin{itemize}
\item An error polynomial = a polynomial was fitted to our rms-error curve (top panel in 
         Fig.~\ref{photErrors}) and given as input.
\item Magnitude limits = our 80\% completeness levels
\item Field of view = ACS field size
\item Galactic coordinates. 
\end{itemize}
The model returns catalogues with the expected number of stars and their V and I magnitudes. 
Generally we find that the number of stars is negligible compared with the number of GCs. 
A more detailed discussion of the effect of this contaminant is given in section 4.1. 

\item Unresolved background galaxies. An estimate of the number of these contaminants 
was made by using the Hubble Deep Field South (\cite{hdfs}). The images are publicly 
available and were observed in the same passbands as our data. Since the HDFS images 
are much deeper than our images, we dimmed the HDFS in order to reach our 80\% completeness 
levels. We did this by dimming the HDFS V and I images with 2.8 and 2.6 magnitudes respectively, 
by multiplying with $10^{-<\Delta mag>/2.5}$. To restore the noise levels of the original images, 
we added Gaussian noise. Next, SExtractor was applied with the same parameters as before, 
selecting objects using the same selection criteria as before and then associating the objects. 
We find 15 sources within $0.55 < V-I < 1.45$, which are evenly spread in V-I. This number, 
together with the number of foreground stars, will be used in Section 4.4 to estimate a 
contaminating background density valid for each galaxy.
\end{enumerate}

\begin{figure*}
    \includegraphics[width=17cm, angle=0]{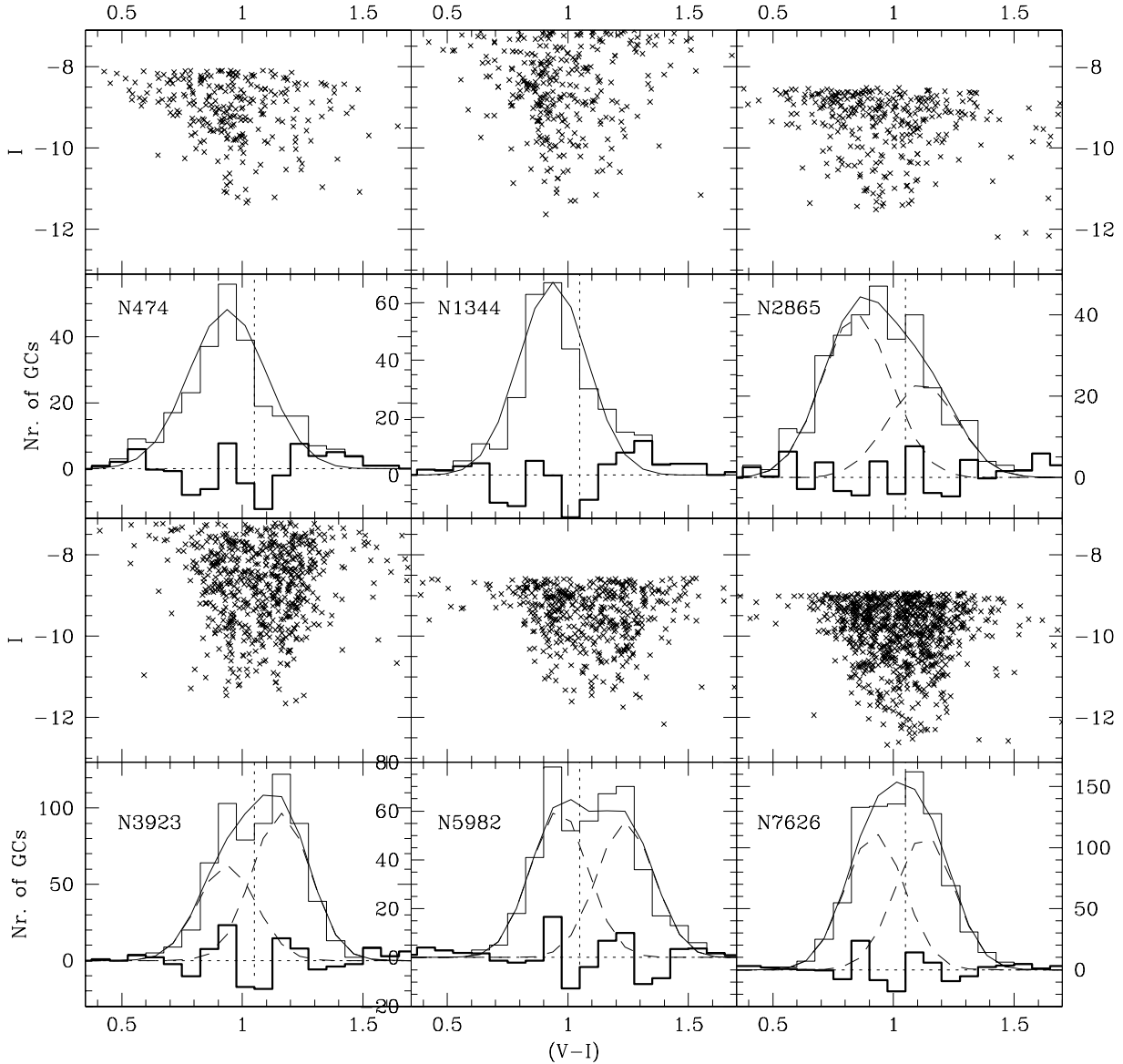}
    \caption{
%Histogram and colour diagram of V-I colours of the globular cluster candidates. 
%The Gaussians drawn, together with a residual histogram (thick line), are fits using the KMM 
%algorithm. All histograms are best fit by two Gaussians rather than one, even for NGC 474 and 
%NGC 1344. Vertical, dashed, lines indicate the separation into red and blue groups. GCSs on the 
%left side of the left vertical line are part of the blue group, GCs on right side of the right 
%vertical line are members of the 'red' group. Furthermore a very shallow histogram is plotted 
%which represents the theoretical number of stars as calculated by the Besan\c con galactic model. 
%NGC 2865 shows a significant number of blue GCs, which is reflected in the value of the blue as 
%returned by KMM of 0.85 (see table 3). NGC 7626: The clump near 1.1 is overluminous by about 1 
%magnitude compared to the other galaxies.
%DC - revised to fit Gerts new plot 6/2/2006
Histograms and colour diagram of V-I colours of the globular cluster candidates. All histograms
are better fit by two Gaussians than one, but for NGC 1344 and NGC 474 we plot only the blue
component as a Gaussian fit, as the numbers in the red peak are not significant. For the other
four galaxies we plot the two Gaussians fit by the KMM algorithm. Residual histograms (thick lines)
represent the difference between the data and the sum of the Gaussian fits. Vertical, dashed lines
represent the separation of the data into red and blue groups at $(V-I)=1.05$. 
}
\label{VIplot}
\end{figure*}

%%GS added new results (slopes blue red +errors in the table
\begin{table*}
\centering
\begin{tabular} {l r r r r r|| r r r r r r r r r}
\hline
\hline
Galaxy& $\mu_{blue}$ & $\mu_{red}$ & ass. bl. & ass. rd. & P-value& log(r) & $\alpha_{red}$ & err & $\alpha_{blue}$ & err & $\alpha_{all}$ & err & $\alpha_{galaxy}$ & err\\
\hline
(1) &    (2)    &   (3)    &    (4)   &   (5)     &   (6)  &   (7)     &  (8)  &    (9)     &  (10)&   (11)    &  (12) & (13) &  (14) & (15) \\
\hline
\object{NGC 474}&  0.90  & 1.22&   224	&34 &	0.035&1.2-2.2   & - & - & - & - &-1.50    &0.14   &-1.87 & 0.03\\
\object{NGC 1344}& 0.92  & 1.21&   268	&49 &	0.004&1.5-2.2   & - & - & - & - &-2.02    &0.17   &-2.23 & 0.02\\
\object{NGC 2865}& 0.85  & 1.12&   192	&111&	0.039&1.35-2.15 &-3.05&0.24& -2.04& 0.17&-2.28    &0.14   &-2.04 & 0.01\\
\object{NGC 3923}& 0.94  & 1.16&   253	&390&	0.036&1.5-2.1   &-0.87&0.15& -0.86& 0.15&-0.90    &0.10   &-1.59 & 0.03\\
\object{NGC 5982}& 0.96  & 1.24&   240	&221&   0.000&1.4-2.1   &-1.98&0.18& -1.22& 0.19&-1.68    &0.13   &-2.23 & 0.04\\
\object{NGC 7626}& 0.92  & 1.13&   455	&448&	0.039&1.5-2.2   &-1.99&0.14& -1.42& 0.13&-1.67    &0.09   &-2.16 & 0.04\\
\hline
\end{tabular}
\label{kmm}
\caption{{\bf{Left}} Output of KMM algorithm (explanation: see Section 4.2) with blue and red V-I peaks (columns 2, 3), KMM assignments of GCs to
blue and red groups (columns 4, 5) and P-value (column 6), See also Fig. 4. {\bf{Right:}} Slopes $\alpha$ 
and errors of the surface densities of red, blue (based on a split at V-I=1.05) and all GCs (columns 8-13), determined applying a weighted 
least squares method on ranges listed in column 7. Last column: slope of galaxy light. Surface densities of 
all, red and blue groups are shown in Figures 8, 9 and 10.}
\end{table*}

\section{V-I distributions and spatial distributions}

\subsection{V-I distributions}

%DC - shortened and clarified this paragraph

To appear in our histograms an object must be detected in both passbands, so to avoid colour bias
in the selection, we selected only objects from our 
source catalogue which are 0.3 magnitudes brighter in both V and I than the respective 80\% completeness limits. 
Figure \ref{completeness} shows that we reach 100\% completeness at these brighter magnitude limits. 
Another advantage of doing this, is that the photometric errors are smaller at these limiting magnitudes 
(0.07 magnitude instead of 0.10). 
The final V-I colour distributions are depicted in the panels of Figure \ref{VIplot}. The upper 
part of each panel shows the distribution of objects in the colour - absolute magnitude plane.
Here, the vertical axis is  
absolute I-band magnitude, which was determined using the distances listed in column 11 of Table 1. 
The lower part shows V-I histograms with binsize 0.075 mag. In these panels the upper, thin-line
histogram represents the data, the lower, thick-line histogram represents the residuals from the 
fits described in Section \ref{sec:bimodality}. 
%DC
%DC - think below is no longer true, so I have commented it out
%DC
%A third, thin-line, histogram shows the expected number of stars in the Besan\c con catalogues; 
%these numbers are very low compared to the number of GCCs and even for NGC 2865, the galaxy at
%the lowest galactic latuitude, will not influence any conclusion drawn from 
%the figures. 

The expected number of stars in the Besan\c con catalogues is very low compared to the number of GCCs and 
will not influence any conclusion drawn from the figures. 
In the NGC 2865 data, the galaxy at the lowest galactic latitude, there is a small bump near V-I=1.6, 
which is also expected from the Besan\c con model. The curves and thick lines are Gaussian fits to the 
histograms and their residuals respectively (see Section \ref{sec:bimodality}). 

\subsection{Components of the colour distributions.}
\label{sec:bimodality}

%DC - rewritten to be much less definite about bimodality, and to place 
%DC - much less reliance on KMM. Should we leave KMM out completely?
%DC - this section is significantly changed - please read it.

Inspection of the histograms in Figure \ref{VIplot} suggests that the
colour distributions of GCs for NGC 3923 and NGC 5982 have the bimodal form
which is normal for bright ellipticals, with blue and red peaks near $(V-I)
= 0.92$ and $(V-I) = 1.18$ respectively. This bimodal form is parameterised
conventionally as the sum of two Gaussians. In section \ref{discussion}
we investigate the ways in which the other four histograms differ from this form.

A check for colour bimodality was made by applying the KMM algorithm (\cite{ashman}) and
DIP test (\cite{gebhardt}) on data points between $0.55 < V-I < 1.45$. KMM was used
in standard mode (fitting two Gaussians with equal $\sigma$'s).
It returns the value P, which indicates if a distribution is better characterised
by a sum of the two Gaussians than a single Gaussian. Table 3 lists the output which
consists of peak values, number counts in each Gaussian and the P-value.

The KMM analysis shows that all of the colour
distributions are better fit by a double than a single Gaussian, but in NGC 1344 and
NGC 474 the numbers of clusters in the red (metal-rich) Gaussian are too small to be 
statistically significant, so we plot in Figure \ref{VIplot} only the blue Gaussian for
these two galaxies. For NGC 7626 and NGC 2865 neither a single nor double Gaussian
provides a good fit to the histogram, and the structure is more complex. However we
do show in Figure \ref{VIplot} the blue and red Gaussians generated by the KMM algorithm.

%%%GS 

The obvious colour bimodality for NGC 3923 and NGC 5982 is confirmed by using the DIP test (\cite{gebhardt}). 
This test calculates the probability of a dip occurring in a supposedly bimodal distribution. 
Applying this test to the colour distributions of our six galaxies gives significant bimodality 
for NGC 3923 and NGC 5982 with the dip probability $P_{dip}=0.99$ and $P_{dip}=0.92$ respectively.

NGC 3923 and NGC 5982 have been studied before. NGC 3923 has been observed
from the ground in the Washington system (\cite{zepf95}). Like us, they found a bimodal
distribution, which, converting their data into V-I (\cite{forbesforte}) and correcting
to our extinction scale, give colours of 1.01 ($\pm{0.05}$) and 1.21 ($\pm{0.05}$),
an offset of +0.06 with respect to our data. Data on NGC 5982 (\cite{kundu})
also revealed a bimodal distribution; applying our extinction scale to their data
gives V-I colours of 0.96 ($\pm{0.03}$) and 1.15 ($\pm{0.03}$). While the blue peak
is the same as ours, their red peak is 0.1 magnitude bluer. We attribute this to
our much larger GC sample.

In the further analysis we
distinguish between red and blue sample by cutting the distribution at $(V-I) = 1.05$,
in order to analyse whether these two populations have different properties. 

%We note that the value of the P value is strongly affected by outliers and 
%is probably not a very reliable measure 
\begin{figure}
    \resizebox{\hsize}{!}{\includegraphics{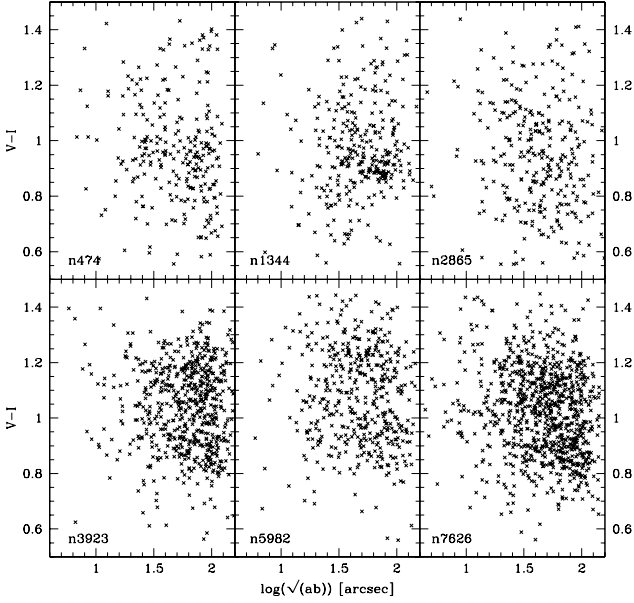}}
    \caption{Globular clusters in V-I (vertical scale) as a function of radius.}
    \label{RVI}
\end{figure}

%OLD
%For NGC 2865 the blue peak is dominant compared with the red and extends further to the
%blue than in other galaxies. The blue peak does not have a simple gaussian form.
%We suggest in section \ref{discussion} that this can be attributed due to a third
%population of young blue clusters, overlapping in colour with the normal old metal-poor
%population, rather than to an unusually populous, broad, and non-guassian (in terms
%of colour distribution) old metal-poor population.

%NEW 
If we fit the V-I histogram of NGC 2865 with 2 Gaussians, as we do for
NGC 3923, NGC 5982 and NGC 7626,
the blue peak extends much further to the blue than in other galaxies. We 
suggest in section \ref{discussion} 
that this can be attributed due to a population of young blue clusters, 
overlapping in colour 
with the normal old metal-poor population.

%OLD
%For NGC 7626 KMM gives maxima at 0.92 and 1.13; the bimodal nature
%was already observed by earlier, but less deep, observations (\cite{kundu}). However
%inspection and the output of KMM show that the simple bimodal form is not a good fit,
%instead the histogram shows a broad flat peak, best explained as a normal bimodal
%distribution, with a clump of bright ($M_I < -11$) GCs with $(V-I) \sim 1.0$ filling
%in the gap. The brighter GCs of NGC 7626 are distributed in several small clumps in
%the colour magnitude diagram, which cannot be explained by random effects: the errors
%in V-I at these magnitudes ($m_V\approx23.0$) are at most 0.04 (Figure \ref{photErrors})
%and smaller than the distances between the clumps. All these clumps add up to
%one big single peak in the histogram, between the normal red and blue populations,
%which illustrates that information is lost if only V-I histograms are analysed.

%NEW
For NGC 7626 KMM gives maxima at 0.92 and 1.13; the bimodal nature
was already observed by earlier, but less deep, observations (\cite{kundu}). However
inspection and the output of KMM show that the simple bimodal form is not a good fit,
instead the histogram shows a broad flat peak. The middle of this peak ( $(V-I) \sim 1.0$) 
is filled by exceptionally bright ($M_I < -11$) and probably young GCs (see Section \ref{sec:recentGCs}).
The colour magnitude diagram shows that these are distributed in several small clumps in, which 
cannot be explained by random effects: the errors 
in V-I at these magnitudes ($m_V\approx23.0$) are at most 0.04 (Figure \ref{photErrors})
and smaller than the distances between the clumps. This galaxy shows that the distribution 
of globular cluster colours is not necessary well represented by 2 Gaussians.

%DC - I don't think there should be a different section here, commented out for the present
%\subsubsection{Results}
%DC - reordered and shortened this section

\begin{figure}
    \resizebox{\hsize}{!}{\includegraphics{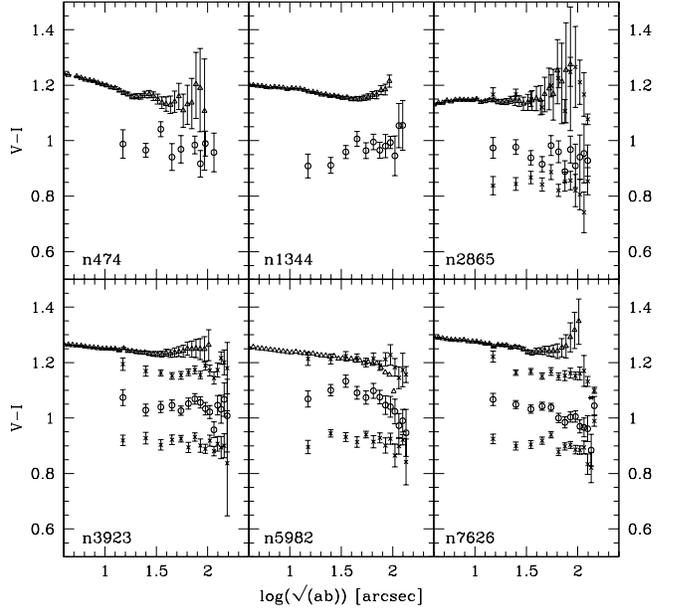}}
    \caption{Average colours of globular clusters (circles: all GCs) and galaxy colour (curve through triangles) as a function of radius. NGC 2865, NGC 3923, NGC 5982 and NGC 7626 also show data points for blue and red groups.}
    \label{RVIave}
\end{figure}

\begin{figure}
    \resizebox{\hsize}{!}{\includegraphics{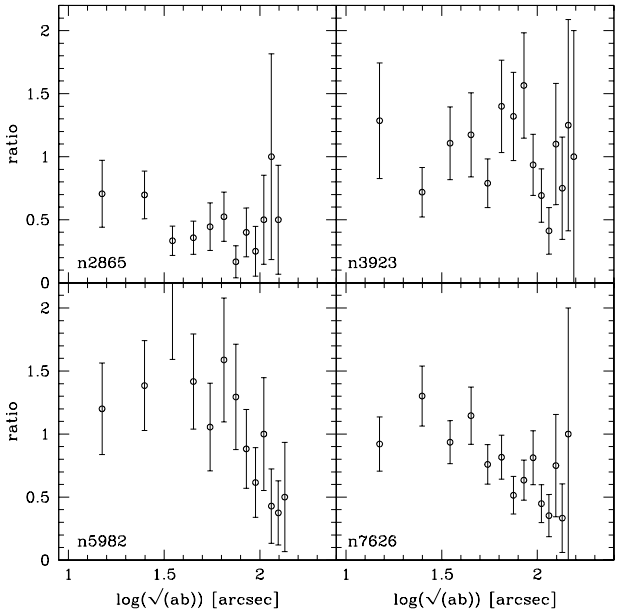}}
    \caption{Ratio of red vs. blue clusters as a function of radius; NGC 5982 and NGC 7626 have strong gradients.}
    \label{GCratio}
\end{figure}

\begin{figure}
    \resizebox{\hsize}{!}{\includegraphics{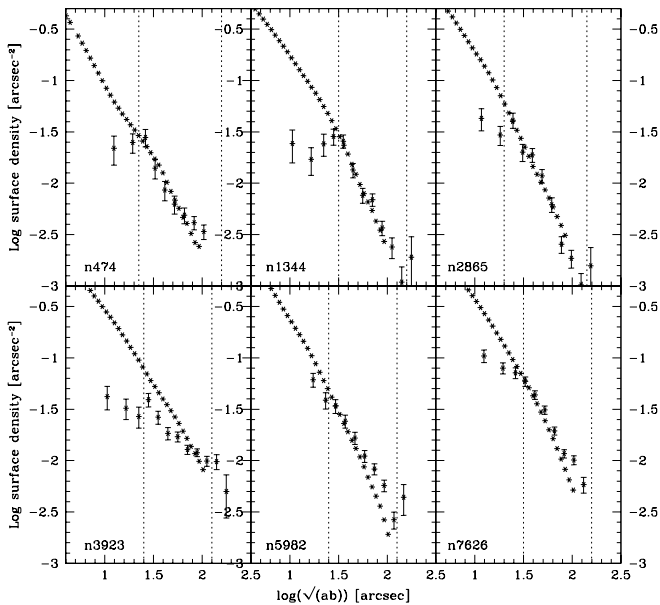}}
    \caption{Surface density of \textit{all} globular clusters, starlike symbols show galaxy surface brightness (arbitrary scale). The slope of the GC surface density was calculated between the dashed vertical lines.}
    \label{SD}
\end{figure}

\subsection{Spatial distributions of the globular clusters}
%DC - minor editing to reflect changes in the previous section

Figure \ref{LOC} shows the spatial distribution of the GCCs. The centers of the galaxies are 
represented by an open square. The distribution of GCs roughly follows the ellipticities of the 
underlying galaxies. The lack of GCs in the central regions is due to dust, bad fits by GALPHOT 
(see Section \ref{sec:observations}) or increased noise. Figure \ref{RVI} shows V-I colours of 
individual GCCs as a function of radius. 
These colours as a function of radius are plotted in bins in Figure \ref{RVIave}. Here, triangles, represent 
the galaxy colour returned by GALPHOT, and circles the average colours of all GCs.  For those galaxies
for which we can separate the population into red and blue groups (Section \ref{sec:bimodality}) we plot as
crosses the mean colours of the red and blue groups respectively. 

Except for NGC 1344, the average colours of all GCs 
tend to be bluer at large radii, confirming earlier results (\cite{forbesfranx}), and reflecting
the colour gradients in the stellar halos. Where there is a significant red subgroup the cluster 
colours in that group match quite well the galaxy colours, except in NGC 3923 where they are 
bluer, as is the case with NGC 1052 (\cite{forbes}). In NGC 5982 and NGC 7626 the strong blue-ward gradient 
in the mean cluster colour is caused by a gradient in the relative fractions of the red and
blue groups, as illustrated by Figure  \ref{GCratio}, where we plot the ratio of the red to 
the blue population against radius.

\subsection{Globular cluster surface densities}
\label{sec:surfacedensity}
%DC - only minor rewording here
%DC - replaced hardwired section and figure references with \refs and \labels

\begin{figure}
    \resizebox{\hsize}{!}{\includegraphics{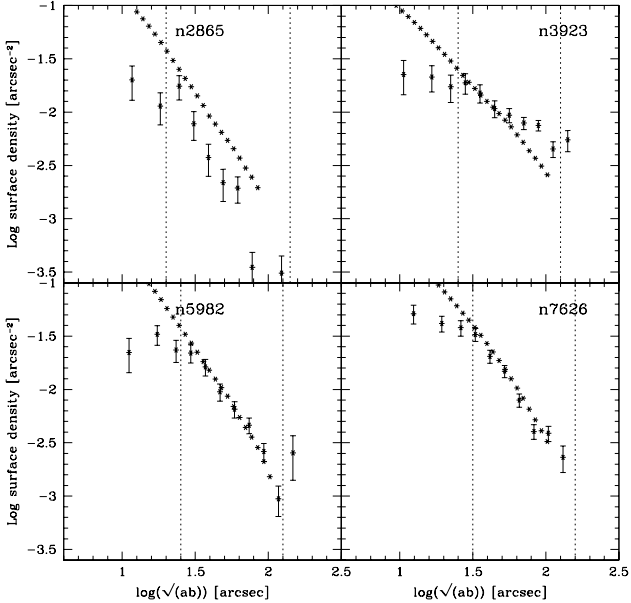}}
%    \includegraphics[height=9cm,width=9cm, angle=0]{ngc7626_3.SDnewt.ps}
%    \resizebox{\hsize}{!}{\includegraphics{NGC 5982_2.SD.ps}}
    \caption{Surface density of \textit{red} globular clusters; starlike symbols show the galaxy light (arbitrary scale). The slope of the GC surface density was calculated between the dashed vertical lines.}
    \label{SDnew}
\end{figure}

\begin{figure}
    \resizebox{\hsize}{!}{\includegraphics{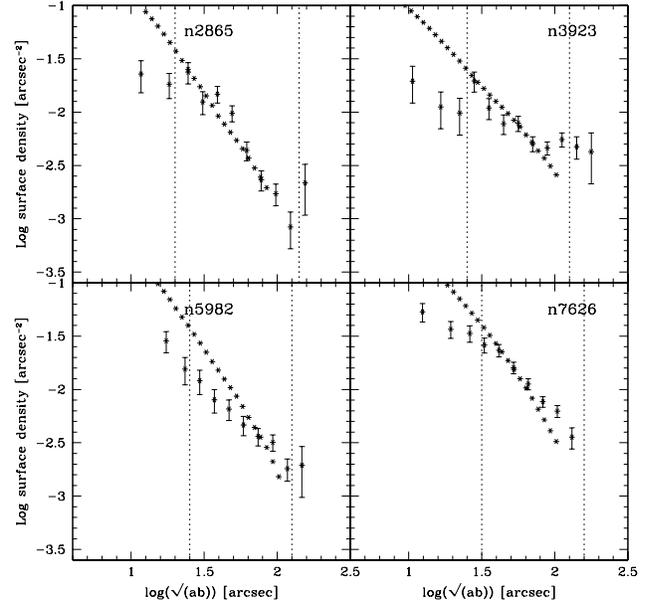}}
%    \includegraphics[height=9cm,width=9cm, angle=0]{ngc7626_3.SDold.ps}
%    \resizebox{\hsize}{!}{\includegraphics{NGC 5982_2.SD.ps}}
    \caption{Surface density of \textit{blue} globular clusters; starlike symbols show the galaxy light (arbitrary scale). The slope of the GC surface density was calculated between the dashed vertical lines.}
    \label{SDold}
\end{figure}

The GC surface densities were calculated in elliptical annuli; their ellipticity was the 
average value in the outer regions as determined by GALPHOT. Dusty regions, bad pixels 
and other bad regions were not used when calculating the effective area of each annulus, from 
which the surface densities are derived. We accounted for contaminating sources by subtracting 
a background density, calculated as described in Section \ref{sec:selection}. These background 
densities were between 4.1 to $8.6 \times 10^{-4}$ arcsec$^{-2}$ for the six galaxies. In 
Figure \ref{SD} we plot the radial surface density of GCs as a function of radius, the 
galaxy surface brightness distribution is also shown in this plot. In Figures \ref{SDnew} 
and \ref{SDold}, we plot the radial density distribution for the clusters in the red group 
and blue group respectively, for those galaxies with a significant red population. 

For all galaxies we note a deficit of clusters in the inner regions, with respect to the 
background light surface brightness distribution. Possible physical causes of this are 
discussed in Section \ref{sec:recentGCs}, but here we consider the possibility of greater incompleteness 
in the inner regions causing this deficit. Two effects could contribute to greater incompleteness 
in the inner regions: confusion due to crowding; and the increased photon noise level in the 
higher surface brightness regions. 

To test the effect of confusion a low density region of 1000 x 1000 pixels, containing 43 
clusters, was cut from the residual image of NGC 7626. As in Section \ref{sec:photometry}, we introduced 
artificial objects at various number densities, using uniform and power-law surface density 
distributions. After using exactly the same detection methods and selection criteria as 
described in Section \ref{sec:selection}, we found that we start losing objects due to confusion effects at 
densities of 0.15 arcsec$^{-2}$, where 5\% are missed. From Figure \ref{SD} we see that even the highest 
density of our galaxy sample (i.e. NGC 7626) is still below this value.

The effect of the galaxy surface brightness on completeness is due to the increased photon 
noise in high surface brightness regions. Within the central 10 arcseconds (13.5 arcseconds 
for NGC 5982) this leads to increased incompleteness and we do not plot points within these 
regions. Outside this, the background photon noise from the galaxy is at most $2\times$ the 
background noise, except for NGC 3923 were the photon noise of the inner data point is about 
$4\times$ the background photon noise. After again introducing artificial objects and using 
the same selection criteria as before, we find that this effect is significant only for NGC 3923: 
the inner two points in Figure \ref{SD} should probably lie somewhat higher.

The radial density distributions of most of the GC systems follow the surface brightness 
distributions of the galaxy light in the outer parts, but show a deficit at small radii 
which we argue is not due to incompleteness due to confusion or photon noise from the 
galaxy. This is typical of ellipticals in general (e.g. \cite{lauer}; \cite{grillmair}; \cite{puzia2}; 
\cite{forbes}; \cite{schweiz} and \cite{brown}). The deficit at the centre is often 
interpreted as evidence for tidal disruption of clusters passing through the core of the 
galaxy (\cite{fallrees}), but to be effective this process requires the clusters to be on 
predominantly radial orbits (\cite{grillmair}; \cite{ostrikerbinney})

NGC 3923 shows a different behaviour, the surface density profile is much shallower than the 
light profile at all radii (already noted by \cite{zepf94}). This is typical of the GC systems of 
{\em some} brightest cluster galaxies, e.g. NGC 4874 (\cite{harris2000}). However it is 
unusual for a galaxy such as NGC 3923, which is the brightest in a small group.

%, because the values for the slopes lie within each others error bars.

Comparison of Figures \ref{SDnew} and \ref{SDold} shows that there are sometimes differences 
between the surface density profiles of red and blue GCs. To quantify these differences we 
fit the linear relation $log(\sigma) = b + \alpha\times r$ by applying 
a linear weighted least squares fit to the outer points of the density curve. 
These points were chosen to lie between an inner and outer radius. The inner radius was defined as the point where the flattening stops, estimated by eye.
The outer radius excludes unreliable data points at larger radii with very low number statistics (usually containing only 1 or 2 objects in the partial elliptical ring located in the very outer corners of the image). We indicate the inner and outer radius in the Figures as vertical dotted lines.

%(with the purpose/thereby to exclude outer points with very low number statistics and hence most affected by incertainties of the background object densities.)

Using the same method and radii we also fitted the galaxy light profile. In columns 7 to 15 
of table 3 we list the radii used and the results of the fits, including errors on the slopes (we used the standard recipe and error formulae, i.e. equations 15.2.6 and 15.2.9 respectively as listed in \cite{numrec})

In NGC 7626, the slope of the red GCs is significantly steeper than the slope of the blue GCs. 
Similar but less significant differences are visible in NGC 2865 and NGC 5982. We checked if these differences 
are due to the choice of our inner radius by shifting the inner radius one data point to the left as well as to the right; we 
found no significant change in slope differences. The slopes calculated for the set of all GCs will 
be used in Section \ref{sec:specificfreq}, where the specific frequency is calculated.

\section{Globular cluster luminosity function in I}
%DC - minor rewording and shortening

In this section, the GCLFs in the I band are 
determined. If the observations are deep enough to cover the absolute turnover 
magnitude (TOM) of $M_{I}=-8.46 \pm 0.03$ (\cite{kundu}), the GCLF can be used as a 
distance estimator. The GCLF can also be used to estimate the number of globular 
clusters in a system; this number then determines the specific frequency $S_N$ (Section \ref{sec:specificfreq}). 
In those galaxies with a significant red population,
we also compare the GCLF of the red and blue samples, which could be different, for instance
if the samples have large M/L differences (\cite{wm}). 

\subsection{Determination}
\begin{figure}
    \resizebox{\hsize}{!}{\includegraphics{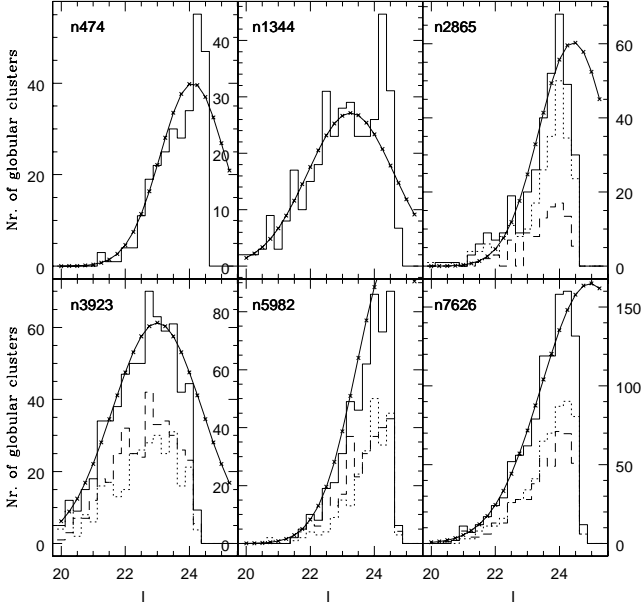}}
%    \includegraphics[height=9cm,width=9cm, angle=0]{ngc7626.I.ICombined3.ps}
%    \resizebox{\hsize}{!}{\includegraphics{ngc5982.V.VOnlyComp.ps}}
    \caption{Globular cluster luminosity functions in the I band. Dashed and dotted 
histograms represent the GCLF of the red and blue groups respectively. Gaussian fits 
are also drawn. Only the GCLFs of NGC 3923 and NGC 1344 extend past the TOM.}
    \label{GCLF}
\end{figure}

We calculated the GCLF in the I band, which is less affected by extinction. 
The GCLFs were constructed from histograms with a binsize of 0.25 magnitudes, 
approximately twice the photometric error of the faintest objects. 
The Besan\c con catalogues were used to correct for contamination by foreground stars. 
The foreground stars (see Section \ref{sec:selection}) were cumulatively subtracted: i.e. if a bin 
contains only three sources, while there are an expected number of five stars, we 
subtracted the remaining two stars from the next bin. We only used objects within 
the 80\% completeness levels and also corrected for incompleteness. The blue and 
red GCLFs were calculated used the red and blue groups defined in Section \ref{sec:bimodality}. 
The GCLFs are shown in Figure \ref{GCLF}, where they are divided in blue and red groups, 
we find no significant differences between the GCLFs of red and blue populations. The peak in the 
GCLF of NGC 1344 at I=24.5 is due to red objects, caused either by photometric errors near
the faint limit, or more likely a local excess in the background. 

%DC - This is a slightly different interpretation of this peak, I think it is more likely
%%%GS: I plotted the surface density of this peak and it follows that of the other, brighter
%%%    GCCs, it would be a coincidence if a local excess would exactly fall on top of the center of this galaxy
%%%

\subsection{GCLF as a distance estimator for NGC 1344 and NGC 3923}
\label{sec:luminosityfunction}
%DC - I have changed the emphasis of this section to globular cluster rather
%DC - than shell galaxy formation. Old version is commented out.
%DC - Also regarding the peak of the TOM, there are competing effects here,
%DC - fading of all the GCs and disruption of the faint ones. Which is stronger?

If the events which formed the shells also formed large numbers of new GCs, then we
might find a difference between the TOM of the GCLF, firstly between red and blue 
populations, and secondly when compared with the canonical value for normal
ellipticals of $M_{I_{peak}} = -8.46$ (\cite{kundu}). However we only cover this 
absolute magnitude for the closest two galaxies, NGC 1344 and NGC 3923. Using
the IRAF tool NGAUSSFIT we find for NGC 3923 $\sigma=1.4 \pm0.1$, $TOM_I = 23.04 \pm0.06$,
and, ignoring the spike in the GCLF at faint
magnitude, for NGC 1344, $\sigma = 1.35 \pm0.10$, $TOM_I = 23.26 \pm0.10$. 

Comparing our GCLF distance estimates for NGC 1344 (m-M=31.72) and NGC 3923 
(m-M=31.50) with the SBF distance moduli (from column 10 in Table 1: 
$31.48 \pm 0.30$ and $31.80 \pm 0.28$ respectively), shows that the 
differences of +0.24 and -0.30 do not exceed the typical errors between 
these two methods (\cite{richtler}). A PNLF distance estimate for NGC 1344 (\cite{theo})
with $m-M=31.40 \pm 0.18$ is in excellent agreement with our value.
In the remainder we will use our calculated distance moduli for NGC 3923 and NGC 1344.

\subsection{Total numbers of GCs}

A Gaussian fit to the GCLF of the other four galaxies is also necessary to 
estimate the specific frequency (Section \ref{sec:specificfreq}). Because the magnitude limit 
is below the TOM of the GCLF, we fit only for $\sigma$ and amplitude, 
keeping the distance modulus fixed within the error bars (0.26 mag.). 
Doing this for NGC 5982 and NGC 7626 give $\sigma$ of 1.5 and 1.25 respectively, and
good fits. For NGC 474 and NGC 2865 we find $\sigma \sim 1.05$, which is unusually
small, and conclude that the calculated $S_N$ will be uncertain, and that deeper data
are required.

%---- CHECK OTHER DISTANCE DETERMINATIONS FOR NGC 474 IN LITERATURE ----

\section{Total number of globular clusters and specific frequencies}
\label{sec:specificfreq}
%DC - only minor rewording here

In this section we estimate the total number of clusters and specific frequency 
$S_N$, which is a measure of the number of GCs per unit galaxy luminosity. 
It is a strong function of galaxy type and environment: late type galaxies 
and isolated galaxies usually have lower $S_N$ than early type and cluster 
galaxies (\cite{harris91}). The calculations are straightforward using the results of previous sections.

The specific frequency is defined by (\cite{hvdb}):

\begin{equation}
S_N = N_{tot}10^{0.4(M_{V}+15)}
\end{equation}

where $N_{tot}$ is the total number of GCs, and $M_V$ is the absolute V magnitude 
of the galaxy, derived from the apparent magnitudes and adopted distances (Table 1). 

The total number of GCs is obtained by using: 

\begin{equation}
N_{tot}=\frac{N_{gauss}}{N_{found}}(N_{missed}+N_{found})
\end{equation}

where $N_{found}$ is the detected number of GCs, corrected for incompleteness, 
$N_{gauss}$ is the expected number of GCs assuming a Gaussian distribution:
\begin{equation}
N_{gauss}=\frac{1}{binsize}A\sigma\sqrt{2\pi}
\end{equation}
with binsize=0.25; amplitude and $\sigma$ are from the Gaussian fits from the previous section .

$N_{missed}$ is the number of GCs expected to be missed due to the fact that the galaxy is much 
larger than the ACS field of view. This number is calculated by extrapolating and integrating the 
GC surface density as determined in Section \ref{sec:surfacedensity} from the outer points 
(listed in column 7 of Table 3), to the point where $\sqrt{ab}$ reaches 50 kpc and 100kpc.
Error sources used in the calculation of $S_N$ are: 
%DC - reformatted list to be on separate lines

1) A photometric measurement error of 0.15 mag.

2) A distance error of $15\%$. This error is taken from \cite{faber} and represents the scatter in the 
$Dn-\sigma$ distance estimator. Comparing our adopted distances with $Dn-\sigma$ distances (see 
columns 11 and 12 in Table 1) shows that this is a reasonable assumption. The distance error was 
applied to both $N_{miss}$ and absolute V.
%DC - question - this will affect both Nmissed and V in the same sense, so will affect
%DC - the ratio less than you might think. Has this been taken account of??
%%%GS: I do not think so; after some analysis it turnes out that the error will be significantly smaller only for NGC 3923 
%%%    

3) The errors in the slope of the GC surface density, listed in column 13 of Table 3.

%DC - reworded and shortened below, plus supporting citations
A possible large error is the unknown value of the slope of the GC surface density outside the 
ACS field of view. We assumed this slope to be equal to the fitted GC density 
profile in the outer part of the galaxies. Especially for our two most nearby galaxies this 
assumption is uncertain; these galaxies are only covered by the ACS to 15-17 kpc, while we 
extrapolate until 50 kpc and even 100 kpc. Data on the slope in the outer regions of 
our most distant galaxy, NGC 7626, and surface density profiles in the literature,
support this assumption (e.g. \cite{rhode2}; \cite{zepf94}; \cite{harris2000}).
Finally, for NGC 3923, wide field data (\cite{zepf94}) show a constant slope to a radius 
of 5.6 arcmin (~34 kpc). Applying a least squares fit to their inner data, covering the ACS fieldsize, 
gives a slope of $-0.82 \pm 0.34$, comparable to our value ($-0.90 \pm 0.10$). Their outer data gives a
slightly steeper slope of roughly $-1.14 \pm 0.12$, which we used to calculate the missed GCs in 
Equation 4.

%We verify this by: 
%1) looking at the slope in the outer regions of our most distant galaxy NGC 7626 and 2) 
%checking the constancy of outer slopes in wide field data in the literature of comparable, 
%field ellipticals. 

%Column 13 of Table 4 shows the range in kpc used to calculate the slope of the density function. 
%Our most distant galaxy NGC 7626 is covered to 37.0 kpc in the ACS field of view. Looking at 
%GC density profile of red and blue clusters in Figures 9 and 10 shows that the slope remains 
%fairly constant in the outer regions. Wide field data (\cite{rhode2}) for three field and cluster 
%ellipticals (coverage until 40, 60 and 100 kpc) also show a constant slope in the outer region. 

Assuming our assumption is justified, we adopt the final values for $S_N$, which are listed in 
columns 6, 7 (50kpc) and 8, 9 (100kpc) of Table 4. There sometimes are large variations between 
$S_N$ calculated at 50 kpc and 100 kpc, this especially true for NGC 3923 and is due to the
shallow slope of the GC density distribution compared with the luminosity distribution. 
\cite{harris98} also noticed such variations and found a ratio 
$S_{N}^{overall}/S_{N}^{40}$ = 1.3, with $S_{N}^{overall}$ the total $S_N$ and $S_{N}^{40}$ 
the $S_N$ within 40kpc. It is important to keep these different definitions of $S_N$ in mind 
if one compares $S_N$ with other authors. Some authors calculate local $S_N$, others extrapolate 
to various values ranging between 25kpc and 200kpc. 

%DC - minor rewording below
Integrated out to 50kpc, we find values of $S_N$ typical for isolated elliptical galaxies,
except for NGC 3923 and NGC 7626, for which we find $S_N = 5.6$ and $S_N = 3.9$ respectively.
These values are more typical of cluster ellipticals, indeed NGC 3923 has the highest $S_N$ of 
any isolated elliptical (\cite{zepf94}). Both of these galaxies are in groups: NGC 3923 
is a dominant group galaxy embedded in an X-ray envelope (\cite{buote}; \cite{pellegrini}); 
and NGC 7626 is the second brightest member of the Pegasus group and is also detected in 
X-rays (\cite{osullivan}).

\begin{table*}
\centering
\begin{tabular} {r r r r r r r r r r r r r r}
\hline
\hline
Galaxy & $N_{found}$ &$N_{missed}$ &error & $N_{gauss}$ & error & $N_{tot}$ & error & $S_{N50}$ & error  & $S_{N100}$ & error & $R_{in}-R_{out}$ (kpc) & $M_V$ (mag)\\
\hline
(1) &    (2)    &   (3)    &    (4)   &   (5)     &   (6)  &   (7)     &  (8)  &    (9)    &   (10) &   (11)     &  (12)  &    (13)    &  (14) \\
\hline
NGC 474 & 284 & 143 & 29 & 405 & 56 & 609 & 94 & 2.1 & 0.5 & 2.7 & 1.0 &  2.5-25.0	&-21.2\\
NGC 1344 & 352 & 93 & 25 & 367 & 28 & 464 & 44 & 1.4 & 0.3 & 1.5 & 0.4  &  3.4-17.0	&-21.3\\
NGC 2865 & 350 & 35 & 6 & 647 & 98 & 712 & 108 & 1.6 & 0.4 & 1.7 & 0.4 &  4.1-25.9	&-21.6\\
NGC 3923 & 660 & 1254 & 162 & 860 & 67 & 2494 & 286 & 5.6 & 1.3 & 8.3 & 3.6  &  3.1-15.4	&-21.6\\
NGC 5982 & 505 & 116 & 18 & 1218 & 161 & 1497 & 202 & 2.6 & 0.6 & 3.0 & 0.9 &  5.1-25.6	&-21.9\\
NGC 7626 & 1051 & 141 & 16 & 2498 & 263 & 2833 & 300 & 3.9 & 0.9 & 4.8 & 1.4 &  7.4-37.0	&-22.2\\
\hline
\label{ntsn}
\end{tabular}
\caption{Column 2: Detected GCs, Columns 3,4: missed GCs + error; Columns 5,6: number 
of GCs + error using the values for the Gaussians determined in section \ref{sec:luminosityfunction}; 
Columns 7,8: total GCs + error applying equation (4); Columns 9,10: $S_N$ within 50 kpc 
with error; Columns 11,12: $S_N$ within 100 kpc with error; Columns 13: ACS range 
where slope has been determined}
\end{table*}

\section{Discussion}
\label{discussion}
In this section a short summary of the properties of 
the shell galaxies is given; a comparison with the GC systems of 'normal' early type galaxies 
is made; the data are compared with predictions of the hierarchical merger scenario simulations
(Beasley et al. 2002); and possible signs for recent GC formation and ages are discussed.

%%%GS changed a reference (merrifield --> ho); ho is the only one known in the literature who classifies 474
%%% 5982 and 7626 as (uncertain) LINERS

\subsection{The shell galaxies}
All of our galaxies are located in low density regions; early type galaxies residing in groups 
and clusters have a much lower probability of  exhibiting shells. (\cite{malincarter}, \cite{col}). 
As well as the presence of the shells, all six galaxies also show visible dust patches and/or lanes, 
mostly in the central regions. 
It has been shown that all galaxies, except for NGC 1344, contain a KDC or show otherwise peculiar 
kinematic behaviour (\cite{hau2}; \cite{hau}; \cite{carter2}; \cite{sauron}; \cite{balcells}). 
NGC 474, NGC 5982 and NGC 7626 may be LINERS (\cite{ho}). Except for NGC 474, unresolved 
X-ray data are available for all galaxies (\cite{osullivan}) and even a 2D X-ray map for NGC 3923 
(\cite{buote}; \cite{pellegrini}). In this paper we assume that NGC 474 and 1344 are E-type galaxies, 
although there is evidence in each case that they night be classified as S0 galaxies.

The environmental and other properties of the galaxies are described briefly below (LGG group numbers 
from \cite{garcia}):\\
\ \\
{\it NGC 474}  -- brightest galaxy of LGG20 (4 members). The galaxy is connected with the small spiral 
NGC 470 via a HI tidal bridge (\cite{schimi2}). Often classified as S0; shows some rotation (\cite{sauron}). 
May be a LINER. \\
{\it NGC 1344} -- located at the outskirts of the Fornax Cluster, $4.9^\circ$ from the central cluster 
galaxy NGC 1399. At this distance the density is comparable to the field galaxy density (\cite{kambas}). 
Sometimes classified as S0; shows rotation within 2 effective radii (\cite{theo})\\
{\it NGC 2865} -- isolated galaxy (\cite{reda}), remains of a rotating HI disk (\cite{schimi}). \cite{hau} 
found a KDC and evidence for a young (0.4-1.7 Gyr) stellar population; two possible explanations were given: 
a starburst or a truncation of the star formation. \\
{\it NGC 3923} -- brightest galaxy of LGG255 (5 members)\\
{\it NGC 5982} -- brightest galaxy of LGG402 (4 members), may be a LINER.\\
{\it NGC 7626} -- second brightest member of the Pegasus group, LGG473, of at least 15 members. This probable 
LINER has a radio-jet, directed NE, and a small HI cloud between 1.5' and 3.0' WSW of the center; the galaxy 
does not contain any HI tidal features (\cite{hibbard}). The core shows orthogonal kinematics to the main 
body (\cite{balcells}), with no emission lines, nor signs of nuclear young populations. A dust lane in 
direction ENE is visible in the inner 15 ACS pixels of the core.\\

\subsection{Comparison of the GC systems with normal ellipticals}
%DC - this paragraph shortened, and citation changed, removed dependence on KMM

%OLD
%Our sample is selected on morphological grounds to have undergone a recent minor
%merger. In this section we investigate the effect of the merger event upon
%the GC systems. All six galaxies have a peak at $0.94 \pm 0.04$ in their
%V-I histograms, in common with 'normal' ellipticals, and interpreted as an old
%metal poor GC population. In four of our six galaxies we see a distribution
%of galaxies redwards of this, this second peak again occurs in the majority of  
%normal ellipticals, and gives more information about the formation history of the
%galaxy.

%NEW
Our sample is selected on morphological grounds to have undergone a recent minor
merger. In this section we investigate the effect of the merger event upon
the GC systems. All six galaxies have a peak at $0.94 \pm 0.04$ in their
V-I histograms, in common with 'normal' ellipticals. In four of our six galaxies we see a distribution
of galaxies red-wards of this, this second peak again occurs in the majority of  
normal ellipticals.

All galaxies show a flattening in the GC density profile near the central regions.
This feature is also a generally seen in other GC systems and is attributed to
disruption processes (\cite{ostrikerbinney}; \cite{pesce}).

%DC - minor rewording of this paragraph
%%%GS include comment of Peletier in the following line.
The properties of the GC systems of NGC 3923 and NGC 5982 are very similar. Both galaxies have blue and 
red populations of comparable size, which peak at roughly the same V-I values near 0.95 and 1.20. Using the 
colours of the GCCs and assuming that the red peak has solar metallicity, the evolution models of \cite{uta}
 (to get the metallicities of the blue clusters) and Fig. 12 from \cite{goudfrooij} indicate that these  
systems are old, $>$ 5 Gyr, and evolved systems. 

NGC 3923 has been extensively studied by \cite{zepf95}. They already noted the very high $S_N$ for 
a galaxy located in a low density environment, which we confirm. \cite{mcl} proposes that high $S_N$ 
galaxies are probably best explained by taking into account the presence of a extended massive X-ray halo. 
However, NGC 3923 is probably in contradiction with these results. While NGC 3923 has a very shallow GC 
surface density profile and also contains a X-ray halo (\cite{buote}, \cite{pellegrini}), its X-ray 
luminosity is about a factor 10 lower (\cite{osullivan}, \cite{fuka}) than the galaxies studied by McLaughlin: M49, 
M87, and NGC1399.

%DC - 
%DC - This is the section where we still have some thinking to do, I have written a paragraph listing
%DC - properties and possibilities as we have been discussing in email exchanges of 8/2/06. But I think
%DC - we need some more thought here
%DC - 
The colour distributions of NGC 474 and NGC 1344 (Figure \ref{VIplot}) are similar, and appear either to be
unimodal, or to have a very low red (metal-rich) component. 
Although KMM returns a combination of a blue and red peak as a good description of the distribution, 
there are very few GCs in the red peak. Such colour distributions are very rare in luminous ellipticals 
(\cite{kundu}). Peng et al. (2006) show that, in early-type Virgo galaxies, the red peak is much more
prominent in the more luminous galaxies. However unimodal, blue, histograms only become common for galaxies 
fainter than $M_V$ = -18 (Figures 4 and 6 of Peng et al.), whereas the absolute magnitudes of NGC 474 and NGC 
1344 are $M_V$ = -21.17 and -21.07 respectively. They are however the lowest luminosity and lowest velocity 
dispersion galaxies in our sample. 

Environmental differences in the early history of the galaxy might lead to differences in the GC (V-I) histograms.
In the merger model, fewer early mergers would mean less red peak clusters. Fewer early mergers might
also cause a galaxy to retain more of its angular momentum, and thus to be an S0 or a rapidly rotating
elliptical. NGC 474 and NGC 1344 both have significant rotation (\cite{sauron} and \cite{theo} respectively),
and both are sometimes classified as S0 galaxies. The
most luminous galaxies in the Peng et al. (2006) Virgo sample with unimodal blue histograms are NGC 4660, an
elliptical with significant rotation (\cite{sauron}); and NGC 4340, an SB0. However there are counterexamples
in the Virgo sample, such as the rotating ellipticals NGC 4564 and NGC 4697 which have red peak dominated colour 
histograms.
Insufficient galaxies have been studied in sufficient depth to analyse any possible correlation between
the rotational properties of ellipticals and their GC colour histograms.
Kundu \& Whitmore (2001b) find a number of unimodal, blue colour histograms among a sample of S0 galaxies
studied with WFPC2 (e.g. NGC 2768), but their sample sizes are small. They do however find that a lower
proportion of their sample of S0s are significantly bimodal, than their equivalent sample of ellipticals
(\cite{kundu}). 

%DC - Comment, should we be citing original measurements rather than Hyperleda???
%%%GS: I took HYPERLEDA values because they weigh-average a number of literature valuesm, which I think is a good method  to indicate the real value.

\begin{figure}
    \resizebox{\hsize}{!}{\includegraphics{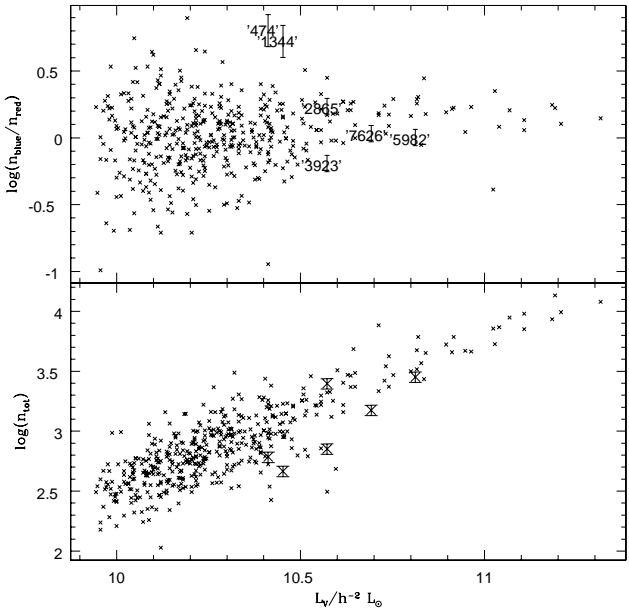}}
    \caption{{\bf{Top:}} Ratio of blue and red clusters vs galaxy luminosity. Simulated data (\cite{beasley}) and the six shell galaxies (NGC numbers). {\bf{Bottom:}}  Total number of clusters vs. luminosity for simulated data and six shell galaxies (large squares)}.
    \label{bmodel}
\end{figure}

Finally we compare these objects with some properties predicted by the hierarchical merger model of \cite{beasley}. 
In these semi-analytical simulations, metal-poor GCs are old and formed in cold gas clumps, 
while the metal-rich are created in time during merger events. Of course, accretion of GCs also takes place during 
the hierarchical build-up of galaxies. These simulations are able to reproduce the many variations in 
the colour distributions of GC systems observed in elliptical galaxies. For instance, in Figure 13 of Beasley et al., a 
blue-peaked V-I distribution is shown which looks similar to our large single peak V-I distributions of NGC 474 
and NGC 1344. His model also roughly reproduces the observed $L_V - N_{tot}$ relation. We plot this relation in 
the lower part of Figure \ref{bmodel} together with our six data points. Our systems seem to have somewhat fewer globular 
clusters than predicted. This is partly explained by the fact that Beasley et al. do not take into account 
disruption processes, which might reduce the number of clusters by 10\%-20\%. The upper 
part of the same Figure shows the logarithm of the ratio between the number of blue and red clusters. NGC 474 
and NGC 1344, the two points with lowest luminosity, lie somewhat outside the point cloud due to their low number 
of red clusters.

\begin{table*}
\centering
\begin{tabular} {r r r r r r}
\hline
\hline
Galaxy & SSP  & $\Sigma$-method & YP/NP ($\Sigma_2$) & Shell Dynamical & GCs \\
\hline
(1) &    (2)    &   (3)    &    (4)    &   (5) & (6) \\
\hline
NGC 474  &$ 7.3^{+2.0}_{-2.4}$ & $5.4 \pm1.9$ & -                 & -        &Old system\\
NGC 1344 & $4.0 \pm1.0 \ddagger    $ & -      & NP (5.5)          & 0.5      &Old system\\
NGC 2865 & Old plus $0.1-1.7 \dagger$& -      & YP (10.6)         & -        &Old plus $0.5-1$\\
NGC 3923 & $2.6^{+0.5}_{-0.6}$ & -            & NP (10.3)         &$0.8-1.3$ &Old system \\
NGC 5982 &$12.3^{+1.9}_{-2.0}$ & $6.8 \pm1.5$ & YP (6.8)          & 0.2      &Old system\\
NGC 7626 &$13.9^{+4.9}_{-2.4}$ & $8.0 \pm1.3$ & NP (1.4)          & -        &Old plus $2-5$\\
\hline
\label{tab: ages}
\end{tabular}
\caption{Age estimates. Column 2: SSP ages from Denicol\'o et al. (2005) except $\dagger$: 
\cite{hau} and $\ddagger$ \cite{kunt}. Column 3: Ages using fine-structure index $\Sigma$ 
method (\cite{schweizseitz}). Column 4: Old (NP) and Young (YP) systems together with their 
$\Sigma_2$ index according to Michard \& Prigniel (2004). Column 5: Dynamical ages of type I shell systems
using Nulsen (1989). Column 6: Approximate age indications from GC (V-I) distribution (this paper)}
\end{table*}

\subsection{Possible evidence for recent GC formation in NGC 7626 and NGC 2865}
\label{sec:recentGCs}

In this section we examine the possibility that the complex (V-I) histograms of NGC 7626 and NGC 2865
provide evidence for recent GC formation, in addition to the old GC population which gives both the
red and blue peaks. In NGC 7626, Figure \ref{VIplot} shows that the brightest GCs of NGC 7626, with 
$M_I\approx -12.5 \pm 0.2$ and V-I=1.1, are more than 1.0 magnitude over-luminous with respect 
to the brightest GCs of the other five galaxies (all at about -11.0), and with respect to its own
universal blue population. It is the most luminous galaxy 
in our sample($M_V = -22.16$), and as such the GCLF would be expected to extend to the brightest 
magnitudes, simply because of the larger population at the sparsely populated faint end
of the GCLF. We compare the GC population of NGC 7626 with that of NGC 4472 (\cite{rhode}), a more luminous
galaxy ($M_V = -22.7$) in the Virgo cluster. NGC 4472 has its brightest clusters in the blue (universal)
peak, at R=19 (equivalent to $M_I = -12.5$). The brightest clusters in NGC 7626 are about 0.2 mag brighter 
and 0.2 mag redder in (V-I) than this, in a less luminous galaxy. This group of bright clusters 
appears in the GCLF of NGC 7626 (Figure \ref{GCLF}) as a small excess near $I = 21$. 

We can compare this population with that found by Whitmore et al. (1997) in NGC 3610, a dynamically
young elliptical. Their Figure 15 illustrates the evolution of a young, metal-rich population
compared with an old, metal-poor population. In time, the young, metal rich, GC population 
will fade in luminosity and will become redder. After three Gyr, this population has become redder 
than the old population but still has several GCs which are brighter than the brightest old metal 
poor GC. This is exactly what is visible in our colour magnitude diagram. Whitmore et al. quantify
the age differences using the $\Delta (V-I)$ vs. $\Delta V_{10}$ diagram (see their Figure 18). 
This diagram represents an age sequence by plotting vertically the difference in V-I between the peaks 
and horizontally the magnitude difference between the 10th brightest globular cluster in the young and
old populations. In NGC 7626 $\Delta V_{10}$ is extremely difficult to estimate, because the young
population is dominated in number not just by the old metal-poor population, but by an old, red, 
metal-rich population as well. Replacing $\Delta V_{10}$ by an estimate of the difference between the 
magnitude of the brightest clusters and the brightest in the blue peak, we estimate a very tentative
merger age of 2 - 5 Gyr. We conclude that NGC 7626 appears to have some young 
GCs, probably created in a recent (2-5 Gyr old) minor merger. These young GCs are superimposed upon a 
much richer, bimodal, old cluster population.

For all galaxies with a substantial red population, we find significantly 
steeper slopes for the red clusters. This is reflected in Figure 
\ref{GCratio}, where gradients are seen in the ratio of red to blue
clusters as a function of position: the red GCs are more centrally
concentrated than the blue GCs. Similar effects are also seen in other
early type galaxies (NGC 1407: \cite{forb06}; NGC 4649 \cite{forbes}; NGC
1399: \cite{dirsch03}; NGC 4636: \cite{dirsch05} and others). 
 
Since the converse situation is never or rarely seen, this must reflect
some important difference between blue and red clusters. This difference
will be related to different formation processes, combined with
disruption, which will affect populations differently depending upon
their orbital structure. Less radial orbits for the red GCs might 
explain this, but without kinematic data on large samples of GCs nothing
conclusive can be said about the cause of these differences.

NGC 2865 shows all signs of a recent merger event: a very luminous shell and a KDC with evidence of a recent, 0.4-1.7 Gyr, 
starburst (\cite{hau}), and an HI disk (\cite{schimi}). 

The (V-I) histogram of the GCs of NGC 2865 (Figure \ref{VIplot}) is more complex than a simple bimodal 
structure, with a population of very blue, not particularly luminous, GCs near (V-I) = 0.7. 
The colour of this population is consistent with an age in the range 0.5 - 1 Gyr (\cite{whit97}), 
consistent with the nuclear starburst age of  0.4 - 1.7 Gyr (\cite{hau}), but the luminosity of the 
brightest clusters is much fainter than predicted. This could be attributed to the small 
number of clusters in the young population, or else to different physical conditions imposing 
a different Globular Cluster Mass Function.

An alternative hypothesis is that the structure in the colour-magnitude
diagrams of the GC systems of NGC 2865 and NGC 7626 is entirely due to
metallicity variations. In the case of NGC 2865 this would require a
population of old clusters with $(V-I) \sim 0.7$, which is too blue for
an old population at any metallicity (e.g. \cite{lee}). In NGC 7626
an old population of intermediate metallicity could fill in the dip in
the CM diagram, but it would have to have a very unusual luminosity function
to produce the numbers of bright clusters that we see.

\subsection{Ages and minor mergers.}

%DC - This section needs to be shortened.
%DC - moreover I have doubts about ALL the SSP ages, in some cases because the populations
%DC - are clearly not SSP (NGC 2865), and in others because I don't think they are based upon 
%DC - very good data (Denicolo).
%DC - Adding dynamical shell age

In this Section we compare ages derived for the stellar population of the
galaxy; for the shell-forming merger events; and for the young GCs that
we argue are present in NGC 2865 and NGC 7626.
Stellar ages derived by comparison with Single Stellar Population (SSP)
models are uncertain for a number of reasons. First, galaxies are clearly
not SSPs, and merger remnants in particular will have at least three episodes
of star formation, corresponding to the two progenitors and to the merger induced
star formation event. Second, the degeneracy between age and metallicity, combined
with uncertainty in isochrone models, and assumptions made about other parameters
of the stellar population, such as the mass function, render SSP ages very
uncertain (\cite{poggi}). Third SSP ages are derived from nuclear spectra only,
which can be unrepresentative of the galaxy as a whole (e.g. \cite{proctor}).
Nevertheless SSP ages have been computed for four of our sample by Denicolo
et al. (2005), for NGC 1344 by Kuntschner et al (2002), and for NGC 2865 a
starburst age has been derived by Hau et al. (1999), and these are listed in Table 5.

Schweizer \& Seitzer (1992) determine a Fine Structure Index, based upon the galaxy morphology, 
and derive an empirical correlation between this and the stellar age. The Fine Structure Parameter 
measures the age since the galaxy as a structure was built up, and this is different from the 
ages of the stars in the galaxy. However it determines a combination of the age of and the 
magnitude of a merger event, so there is a degeneracy here as well. Schweizer \& Seitzer quote 
Fine Structure ages for three of our sample which are listed in column 3 of Table 5. A similar 
approach was carried out by Michard \& Prugniel (2004). They divided their peculiar elliptical galaxy sample 
into a normal (NP), reddish, sample, with no signs of a young stellar population and a bluish 
sample (YP) with evidence for a younger stellar population mixed with an old one. They list 
five of our galaxies and their results are listed in the fourth column of Table 5.

Dynamical ages for phase-wrapped, type I shell systems such as NGC 1344, NGC
3923 and NGC 5982 can be derived from shell radii and spacings of the
outer shells, where dynamical friction and tidal effects are
unimportant in determining the particle distribution (\cite{nulsen}). We
used Nulsen's equation 88 to derive the dynamical ages. Mass estimates were 
taken from the literature (NGC 1344: \cite{theo}; NGC
3923: \cite{fuka}; fundamental plane masses: eq. 7 of \cite{vds}). The
calculated ages are shown in Table 5 and are of the order of 
several 100 Myrs. It is clear that these young ages are not reflected in
the GC populations for these galaxies, implying that the shell forming
event did not in these cases give rise to a substantial GC population.

%DC - Its possible we can use Nulsen's equation 88 to produce dynamical age estimates
%DC - for NGC 1344 and NGC 5982, I will look.

%DC - below shortened considerably, I don't think we can say very much here.

Our sample consists of minor merger remnants, and only if it is possible to separate the 
old and merger related populations, as in the case of NGC 2865, can we see a correspondence
between the different age estimators. More detailed studies of the stellar populations of
all of the other galaxies would be valuable, as would better theoretical estimates of the 
dynamical ages of the more complicated Prieur (1990) Type II and III shell systems.

\section{Conclusions}
%DC - shortened, and made some conclusions less definite

The properties of the globular cluster systems of six shell galaxies were analysed.
For NGC 2865 and NGC 7626 we observe anomalous features in the (V-I) histograms, in 
addition to the bimodal structure which is normal for ellipticals. The features
represent excesses at intermediate colour in NGC 7626, and at very blue colours
in NGC 2865, consistent with a small population of GCs, of age 2-5 Gyr and 0.5-1 Gyr
respectively, possibly formed in the merger event which created the shells. In each case
the young population is dominated by the much larger, old, bimodal distribution. 

The data for two galaxies (NGC 1344 and NGC 3923) allow the determination of their globular 
cluster luminosity function distances. Fitting Gaussians to their GCLF give distances moduli 
of 31.72 and 31.50 for NGC 1344 and NGC 3923 respectively.

%DC - minor rewording and citations in the next paragraph
The properties of NGC 3923 and NGC 5982 are very similar. Their bimodal V-I distributions and 
radial density profiles of blue and red clusters are typical for old GC systems in ellipticals. 
NGC 474 and NGC 1344  show one single blue peak and very shallow red peaks in their V-I histograms. 
These properties are unusual for bright ellipticals (\cite{kundu}, \cite{peng06}) and may indicate 
less early mergers in their formation history.

NGC 3923 and NGC 7626 have higher specific 
frequencies (respectively 5.6 and 3.9 within 50kpc) than normal for galaxies located in a low 
density environment. The luminosity of the X-ray halo detected in the former galaxy is probably 
not sufficient to explain its high $S_N$ as proposed by \cite{mcl}. The $S_N$ of the other 
galaxies have values of around 2 and are typical for galaxies located in a low density environment. 

Although for some of these six galaxies new GCs may have been formed recently, the general 
properties (like V-I distributions and flattening of GC density profile) of the globular 
cluster systems of these shell galaxies do not deviate systematically from 'normal' elliptical galaxies. 

In a future paper (Sikkema et al., in preparation) we will investigate the morphology and stellar 
populations of the diffuse galaxy, in particular in the shells.

\section {ACKNOWLEDGEMENTS}

We thank an anonymous referee for helpful comments. The ACS observations 
were made with the NASA/ESA Hubble Space Telescope, operated by the Space Telescope Science Institute, 
which is operated by the Association of Universities for Research in Astronomy under NASA contract
NAS 5-26555. These observations are associated with programme GO9399.

\listofobjects
\end{document}